%% file: sfs.tex
\documentclass[prl,twocolumn,superscriptaddress,showpacs,amsmath,amssymb,hidelinks]{revtex4-1}

\usepackage{hyperref}
\usepackage{graphicx,xcolor}
\usepackage{amsmath}
\usepackage{amssymb,bm,bbm}
\usepackage{esint}

\DeclareMathOperator{\Tr}{Tr}
\DeclareMathOperator{\tr}{tr}

\DeclareMathOperator{\sgn}{sgn}
\renewcommand{\vec}[1]{\bm{#1}}

\definecolor{MS-color}{RGB}{128,0,128}

\definecolor{PV-color}{rgb}{0.97,0.57,0.11}

\begin{document}

\title{Quasiclassical expressions for the free energy of  superconducting systems}

\date{\today}

\author{Pauli Virtanen}
 \affiliation{Department of
Physics and Nanoscience Center, University of Jyv\"askyl\"a, P.O.
Box 35 (YFL), FI-40014 University of Jyv\"askyl\"a, Finland}

\author{Artjom Vargunin}
\affiliation{Department of
Physics and Nanoscience Center, University of Jyv\"askyl\"a, P.O.
Box 35 (YFL), FI-40014 University of Jyv\"askyl\"a, Finland}
\affiliation{Institute of Physics, University of Tartu, Tartu, EE-50411, Estonia}

\author{Mikhail Silaev}
 \affiliation{Department of
Physics and Nanoscience Center, University of Jyv\"askyl\"a, P.O.
Box 35 (YFL), FI-40014 University of Jyv\"askyl\"a, Finland}
\affiliation{Moscow Institute of Physics and Technology, Dolgoprudny, 141700 Russia}

\begin{abstract}
In the seminal work by G. Eilenberger, Z. Phys. {\bf 214}, 195 (1968) the quasiclassical expression for the free energy 
  of spin-singlet superconductor
   has been suggested.
 Starting from the
  Luttinger-Ward formulation we derive the Eilenberger free energy and find its generalization for superconductor or superfluid  with spin-triplet correlations.
  Besides ordinary superconductors with various scattering mechanisms, the obtained free energy functional can be used for systems with spin-triplet pairing 
  such as superfluid $^3$He and superconducting systems with
  spatially-inhomogeneous exchange field or spin-orbit
  coupling.  Using this general result we derive the simplified expression for the free energy
  in the diffusive limit in terms of the momentum-averaged propagators. 
\end{abstract}

\maketitle


Quasiclassical approximation is one of the basic tools in the theory of Fermi systems. It is based on the separation of scales when the characteristic wavenumbers and frequencies of interest are much smaller than the Fermi wave vector and energy. In the field of superconductivity the quasiclassical approach has been introduced in the classical works  \cite{Eilenberger1968, Larkin1968}. This technique has been applied for  various systems. 
The prominent examples are the Usadel theory for dirty superconductors \cite{PhysRevLett.25.507},  microscopic description of superfluid $^3$He  \cite{rainer1976,serene1983-qat,vorontsov2003-tpt}, theories of superconducting hybrid structures \cite{RevModPhys.77.1321,
RevModPhys.77.935} transport properties of mesoscopic superconducting devices \cite{Belzig1999} and superconductors with spin-splitting field \cite{RevModPhys.90.041001}.

An important component partially missing in previously developed quasiclassical theories is a convenient expression for the free energy that would not involve the complication of a $\lambda$-integration procedure over the general coupling constant \cite{rainer1976,serene1983-qat,thuneberg1984-efp} or additional limiting approximations. Such an expression has been introduced by Eilenberger \cite{Eilenberger1968} for the particular case of spin-singlet superconductors
where the correlation functions have trivial spin structure.
Different forms of variational functionals yielding the quasiclassical equations as their saddle points have also been discussed in the framework of  non-linear $\sigma$-models \cite{muzykantskii1995,andreev1996,altland2000-ftm,PhysRevB.64.014512}. Although the expression by Eilenberger has been used in many subsequent works, its relation with the general Luttinger--Ward free energy functional \cite{PhysRev.118.1417,serene1983-qat} or the variational functionals does not appear to have been explicitly clarified. Furthermore, its generalizations to systems with spin-triplet superconducting correlations have not been discussed in detail. In the present Letter, we resolve these issues by evaluating the $\lambda$-integral analytically, and obtain free energy functionals for general spin structure.
We demonstrate that different versions of the free energy discussed in the previous works \cite{Eilenberger1968,serene1983-qat,muzykantskii1995,andreev1996,kusunose2004-qts} coincide with
the Eilenberger-type expression rigorously derived from the general Luttinger-Ward functional \cite{PhysRev.118.1417}.

{\bf General formulation.}
General expression for the free energy of a many-body fermionic system has been derived by Luttinger and Ward \cite{PhysRev.118.1417}. Later this expression has been adopted by Serene and Rainer \cite{serene1983-qat}
to describe the superfluidity of a  Fermi liquid
using the expansion in small parameters determined by the ratio of pairing energy to the Fermi energy. 
The same approach works for the BCS model of superconductivity in metals.
This expansion is formulated in terms of the quasiclassical propagator \cite{Eilenberger1968}
 \begin{align} \label{Eq:QuasiclassicalG}
     \hat g =  \frac{i}{\pi} \fint d\xi_p  \hat\tau_3\hat G 
 \end{align}
 where $\hat G (\bm r, \bm p,  \omega)$
 is the exact Green's function and $\xi_p = p^2/2m - E_F$ is the kinetic energy of electrons relative to the Fermi level. 
 The quasiclassical Green's function $\hat {g} ({\bm n}_p, {\bm r}, \omega)$ 
 is a
$4\times 4$ matrix in a combined spin and Gor'kov-Nambu space and 
depends on the
direction of quasiparticle momentum ${\bm n}_p = \bm p/p$, the position in real space $\vec{r}$
and the Matsubara frequency $\omega$.

Integration in (\ref{Eq:QuasiclassicalG}) is implemented in the vicinity of the Fermi sphere and the off-shell
contribution is neglected  
resulting in the following expression for the free energy \cite{serene1983-qat}
  \begin{align} \label{Eq:OmegaQgen}
      \Omega = \frac{1}{2} {\rm Tr} [\hat\Sigma \hat g - \frac{1}{\pi}\int d\xi_p \ln (-i\hat\Sigma  - \hat G_0^{-1})] + \Phi [\hat g]
  \end{align}
  where $\hat \Sigma$ is the self-energy and the last term is the functional generating the self-energy   $\hat\Sigma = - 2\delta \Phi/\delta \hat g^T$.
  The generalized trace operator in Eq.~\eqref{Eq:OmegaQgen} defined as 
$\Tr=\pi T N_0\sum_{\omega_n} \int \frac{d\Omega_p}{4\pi}\tr$ contains a Matsubara sum, Nambu
and spin traces, integration over $\bm n_p$  directions, and the density of states at the Fermi level $N_0$.
The superconducting pairing is determined by a
contribution to the generating functional in~\eqref{Eq:OmegaQgen},
$\Phi_\Delta[\hat{g}]=  - \Tr ( \hat \Delta[\hat g] \hat g)/4$,
where $\hat \Delta = \hat \Delta [\hat g]$ is given by the  self-consistency relation for the gap function,
which is a linear functional that describes all possible types of pairing. 
In addition, there are other contributions to $\Phi$, e.g., from various scattering mechanisms,
including potential impurity scattering, spin-orbital and spin-flip relaxation \cite{1902.09297}.

The operator $\hat G_0^{-1} = i(\omega \hat\tau_3 + \bm v_F\cdot\hat\nabla) -\hat V$ contains a spatial derivative  in the direction determined by the  Fermi velocity $\bm v_F =  v_F\bm n_p$ and the spin-dependent potential energy  $\hat V= \hat V (\bm r)$. 
Therefore  calculation of the logarithmic term in 
(\ref{Eq:OmegaQgen}) is rather nontrivial. One way to do this is based on the observation (c.f. \cite{thuneberg1984-efp}) $ \int d\xi_p \partial_\lambda \Tr\ln (-i\lambda\hat\Sigma - \hat G_0^{-1}) = \pi\Tr \hat\Sigma \hat g_\lambda $
 resulting in the general expression for the free energy density of a non-uniform superconductor or Fermi superfluid
\cite{serene1983-qat,thuneberg1984-efp,vorontsov2003-tpt,rainer1976}:
\begin{align}
  \label{eq:origina-qcl}
  \Omega[\hat{g},\hat{\Sigma}]
  &=
  \frac{1}{2}\int_0^1 d{\lambda} \Tr [
    \hat{\Sigma}( \hat{g} - \hat{g}_\lambda)]
  + \Phi[\hat{g}]
  \,,
  \\ 
  0 &= \vec{v}_F \cdot \check{\nabla} \hat g_\lambda + [\hat M_\lambda, \hat{g}_\lambda]
  \,,
  \quad
  \hat{g}_\lambda^2 = 1
  \,,
  \label{eq:glambda-condition}
\end{align}
where the normal-state part has been subtracted from $\Omega$, $\Phi$, and $\hat{\Sigma}$.
We denote $\hat M_\lambda =\hat{\Lambda} + \lambda\hat{\Sigma}$ and
$\hat{\Lambda}=(\omega + i
\hat V )\hat{\tau}_3$. 
Here, $\hat{g}_\lambda=\hat{g}_\lambda[\hat{\Sigma}]$ is a functional of the variational self-energy,
which gives the quasiclassical Green function (GF).
It satisfies the Eilenberger equation and the normalization condition (\ref{eq:glambda-condition}).
The potential energy can include a Zeeman  term
$\hat V = {\bm \sigma}\cdot \bm h$ with a general texture of exchange field $\bm h = \bm h(\bm r)$ as well as spin-orbit coupling (SOC). 
The latter however is more conveniently included in the covariant differential operator defined as 
  ${ {\check \nabla}_k}  = \nabla_k - i e [ \cdot , \hat\tau_3  A_k ] - i [ \cdot , {\cal A}_k ]  $, where $A_k$ are the components of the the vector potential and ${\cal A}_k ={\cal A}_{kj}\sigma_j $
is the SU(2) gauge field for the SOC.

Expression  (\ref{eq:origina-qcl}) can be used  for any weakly-coupled 
superconducting or superfluid state with arbitrary  pairing interactions
and fields  $\bm A (\bm r)$,  
$\bm h (\bm r)$ and $\hat{\mathcal{ \bm A}}(\vec{r})$. However,
the remaining $\lambda$-integration necessitates solving Eq.~\eqref{eq:glambda-condition} for
the auxiliary propagator $\hat g_\lambda$ for many $\lambda$. This makes the functional (\ref{eq:origina-qcl}) less convenient for numerical work, and hinders analytical  calculations in certain limiting cases such as e.g. in the dirty limit with small impurity scattering time $\tau$ or in the Ginzburg-Landau regime close to the critical temperature.

A simpler free-energy functional without $\lambda$-integration  
has been suggested by Eilenberger \cite{Eilenberger1968} for the particular case of
spin-singlet superconductor and in the absence of spin-rotating fields (i.e.
collinear $\bm h$ and $\mathcal{\bm A} =0$), but without a systematical procedure for
extending the result beyond this case. Below, we discuss a way to extend it.

{\bf $\lambda$-integration.}
The $\lambda$-integral in \eqref{Eq:OmegaQgen} can be evaluated using an approach suggested in Ref.~\onlinecite{1706.08245}. 
Let us note the general relation 
\begin{align}
  \label{eq:gsigmaterm}
  \Tr[\hat{\Sigma}(\hat{g} - \hat{g}_\lambda)]
  =
  \partial_\lambda
   \Tr[\hat M_\lambda (\hat{g} - \hat{g}_\lambda)] 
  + \Tr [ \hat M_\lambda \partial_\lambda \hat g_\lambda]
  \,.
\end{align}
Here the first term on the r.h.s. is a full $\lambda$-derivative and easily integrated, but
further treatment is needed for the second term. 
To calculate its contribution,  we note that the variation of GF
preserving the normalization condition $\hat g^2 =1$ can in general be
written as $\delta g = [\delta \hat W, \hat g] $ where 
$\delta \hat W$ is a matrix with infinitesimal coefficients. Hence,
the derivative can be represented as
\begin{align}
  \partial_\lambda \hat g_\lambda = [ \hat W_\lambda, \hat g_\lambda]
  \,.
\end{align}
Using Eq.~\eqref{eq:glambda-condition}, the last term in Eq.~\eqref{eq:gsigmaterm} can be written as
\begin{align}
  \label{eq:Adgvar}
  \Tr [ \hat M_\lambda \partial_\lambda \hat g_\lambda] =
  \Tr [
       ( \vec{v}_F\cdot\check{\nabla} g_\lambda) 
       \hat W_\lambda ] .
\end{align}
To proceed, let us now assume that there exist a functional density $E[\hat{g}]$
whose variation over the GF components yields the gradient term 
\begin{align} \label{Eq:ElVarialtion}
  \delta
  \int d^3r\,
  E[\hat{g}]
  =
  \int d^3r\,
  \Tr [ (\vec{v}_F\cdot \check{\nabla} \hat g) \delta \hat W]
\end{align}
Then from Eq.~\eqref{eq:Adgvar} we get
\begin{align}
  \label{Eq:ElVariationIntegral}
  \int d^3r\,
  \Tr [ \hat M_\lambda \partial_\lambda \hat g_\lambda]
  =
  \frac{d}{d\lambda}
  \int d^3r\,
  E[\hat{g}_\lambda].
\end{align}
Finally, we can perform the $\lambda$ integration to 
obtain the general expression for the free energy functional:
\begin{align} \label{Eq:IntegratedOmega}
  \Omega[\hat{g},\hat{\Sigma}]
  &=
  \frac{1}{2} E[\hat{g}_1[\hat{\Sigma}]]
  +
  \Phi[\hat g]
  +
  \frac{1}{2} \Tr \bigl[\hat{\Lambda} (\hat g_n - \hat{g}) \bigr]
  \\
  \notag
  &\quad + \frac{1}{2} \Tr[(\hat{\Lambda}+\hat{\Sigma})(\hat{g} - \hat{g}_1[\hat{\Sigma}])]
  \,.
\end{align}
where $\hat{g}_n\equiv\hat{g}_{\lambda=0}=\sgn(\omega)\hat\sigma_0\hat{\tau}_3$ and we have
chosen $E[\hat{g}_n]=0$.  Using
Eqs.~\eqref{eq:glambda-condition},\eqref{Eq:ElVarialtion}, the
saddle-point equations $(\delta/\delta\hat{g})\Omega=0$,
$(\delta/\delta\hat{\Sigma})\Omega=0$ can be reduced to
$\hat{\Sigma}_*=-2(\delta/\delta\hat{g}^T)\Phi$ and
$\hat{g}_*=\hat{g}_1[\hat{\Sigma}]$, which indeed correspond to the
quasiclassical equations.

The value of the functional at the saddle point gives the free energy:
\begin{align}
  \label{Eq:substitutedF}
  \Omega
  =
  \frac{1}{2} E[\hat{g}_*]
  +
  \Phi[\hat{g}_*]
  +
  \frac{1}{2} \Tr \bigl[\hat{\Lambda} (\hat g_n - \hat{g}_*) \bigr]
  \,.
\end{align}
The gradient functional $E[\hat g]$ remains to be determined.

{\bf The functional $E$.}
In spin-diagonal systems the gradient
terms of the expression given by Eilenberger~\cite{Eilenberger1968}
constitute $E[\hat{g}]$. In the presence of general spin-triplet correlations, the situation is
more complicated, and we need to find a functional satisfying  Eq.~\eqref{Eq:ElVarialtion}.

Let us first state the result:
\begin{align}
  \label{eq:Egeneral}
  E[\hat{g}]
  &=
  \frac{1}{2}
  \Tr  (
  \hat{g}[\hat{\tau}_t,\hat{g}]
  \vec{v}_F\cdot\check{\nabla}
  [\hat{\tau}_t,\hat{g}]^{-1}
  )
  \,,
\end{align}
where $\hat{\tau}_t$ is an arbitrary matrix field normalized to $\hat{\tau}_t^2=1$.  In
the singlet case, we can denote $\hat\tau_t = \hat{\bm \tau} \cdot \bm
t$ where $\hat{\bm \tau} =(\hat\tau_1,\hat\tau_2,\hat\tau_3)$ and $\bm
t=(t_x,t_y,t_z)$ a vector normalized as $t^2=1$. The field can be inhomogeneous in space.
Indeed, using the properties $\hat{g}^2=1$, $\hat\tau_t^2=1$, and
$\delta \hat{g}=[\delta \hat W,\hat{g}]$, a straightforward
calculation \cite{SM} yields the variation~\eqref{Eq:ElVarialtion}
for any texture $\hat{\tau}_t(\bm r)$. The gradient functional is not
unique.

The above functional can be found as follows: we first express the
Green function in terms of  Riccati parameters \cite{schopohl1995-qsa,*cond-mat/9804064,eschrig2000-dfi,Sauls2009} $a$,
$b$ which are $2\times 2$ matrices in spin space, and
\begin{equation}\label{Ricatti-g}
  \hat {g} = \begin{pmatrix}
    (1-a b )^{-1} & 0 \\
    0 & (1-ba)^{-1} \\
  \end{pmatrix}
  \begin{pmatrix}
    1+a b & 2 a \\
    -2 b & -1 - b a \\
  \end{pmatrix}
  \,.
\end{equation}
This form automatically satisfies  the normalization condition $\hat{g}^2=1$. Moreover, the
Eilenberger equations (\ref{eq:glambda-condition}) imply that $\hat a$, $\hat b$ obey Riccati
equations \cite{cond-mat/9804064, Sauls2009}
\begin{align}
\label{eq:eilenberger1}
 \bm v_F \cdot \check{\nabla} a
 & - (2\omega +  a{\bar\Delta}) a +  {\Delta} = 0 ,
 \\
 \label{eq:eilenberger2}
 \bm v_F \cdot \check{\nabla} b
 & + (2\omega -  b {\Delta}) b + {\bar\Delta} = 0,
\end{align}
It is relatively straightforward to find an Ansatz
functional that has Riccati equations as its saddle point.  For
example, one can use the functional (\ref{Eq:substitutedF})
with \cite{virtanen2019-req,SM}
 \begin{align} \label{Eq:RiccatiFreeEnergy}
 E= {\rm Tr}
  [ ( a^{-1} -  b )
  (\vec{v}_F\cdot\check{\nabla} )
  ( a^{-1} +  b)^{-1} ]
  \end{align}  
Rewriting (\ref{Eq:RiccatiFreeEnergy})   in a parametrization-independent way yields Eq.~\eqref{eq:Egeneral}
with $\hat \tau_t =\hat\tau_3$.
To obtain the free energy in a form similar to that suggested by Eilenberger we 
can consider Nambu components of the quasiclassical propagator
$\hat g = (g, f; \bar f, \bar g)$
where the normal $g, \bar g$ and anomalous $f, \bar f$
components are $2\times 2$ matrices in spin space.
The the general form (\ref{eq:Egeneral}) with  $\hat \tau_t =\hat\tau_3$ yields
\begin{align}
  E  =
  \frac{1}{2}
  \Tr  \left[
    gf(\bm v_F\cdot\nabla) f^{-1} + \bar{g}\bar{f}
    (\bm v_F\cdot\nabla)
    \bar{f}^{-1}
    \right]\,,
\end{align}
which clearly reduces to Eilenberger's result in the spin-diagonal case.

The expression~\eqref{eq:Egeneral} is not defined at points where
$[\hat{\tau}_t,\hat{g}_\lambda]$ is not invertible. Such points, if they
occur inside the region swept by the $\lambda{}$-integration,
produce imaginary winding number contributions. \cite{SM} For example, in the singlet case, $E[\hat{g}_n]\hat{=}i
{\vec{v}_F\cdot\nabla\psi}$ (excluding the Matsubara sum and angle
average), where $\psi$ is the polar angle of rotation of the unit
vector $\vec{t}$ around the $z$-axis.  Since the free energy is
real-valued, such contributions are removed by taking the
real part.
Moreover, in practice, one should  choose $\hat{\tau}$ to avoid
singularities in $E[\hat{g}_{\lambda=1}]$. Close to normal state where
$\hat{g}\approx\hat{\tau}_3$, $\hat{\tau}_1$ is a stable choice.
Alternatively, given a decomposition
$\hat{g}_0(x)=U_0(x)^{-1}\hat{\tau}_3U_0(x)$ for some fixed
$\hat{g}_0(x)\approx\hat{g}(x)$, one can choose
$\hat{\tau}_t(x)=U_0(x)^{-1}\hat{\tau}_1U_0(x)$.  This is also
applicable in the spin-diagonal problem.

Writing $\hat{g}=\hat{U}\hat{\tau}_3\hat{U}^{-1}$ we can also recognize $\hat{W}_\lambda=(\partial_\lambda{}\hat{U})\hat{U}^{-1}$ so that
\begin{align} \label{Eq:TopologicaInvariant}
  \int_{M}d\lambda\,ds\,\tr[\partial \hat{g}_\lambda \hat{W}_\lambda]
  =
  \int_{M}
  d(E_sds + E_\lambda d\lambda)
  \,,
\end{align}
where $M=[0,1]\times[-\infty,\infty]$,
$E_s=-\tr[\hat{\tau}_3\hat{U}_\lambda^{-1}\partial{}\hat{U}_\lambda]$,
$E_\lambda=-\tr[\hat{\tau}_3\hat{U}_\lambda^{-1}\partial_\lambda{}\hat{U}_\lambda]$,
and $\partial=\vec{n}_p\cdot\check{\nabla}$ is the long derivative
vs. the coordinate $s$ along the quasiclassical
trajectory. Hence, the gradient term can also be expressed as a Berry/Wess--Zumino term \cite{witten1984}
associated with the quasiclassical Green function. A kinetic term of this type
was obtained in
Refs.~\cite{muzykantskii1995,andreev1996} for the
action of the ballistic $\sigma$-model, which is closely related to
the present problem.

Finally, to compute the term in Eq.~\eqref{Eq:IntegratedOmega},
we can substitute $\check{\nabla}\hat{g}$ from Eq.~\eqref{eq:glambda-condition}
into Eq.~\eqref{eq:Egeneral}. Direct calculation gives (for $\check{\nabla}\hat{\tau}_t=0$),
\begin{align}
  \Omega
  &=
  \frac{1}{2} \Tr (
  \hat{\Sigma}\hat{g}
  +
  \hat{\Lambda} \hat g_n
    -
    [\hat{\tau}_t,\hat{\Lambda}+\hat{\Sigma}][\hat{\tau}_t,\hat{g}_1[\hat{\Sigma}]]^{-1}
    )
    +
  \Phi[\hat g]
  \,,
\end{align}
whose real part is equal to Eq.~\eqref{eq:origina-qcl}, if integrated over
space.


{\bf Diffusive limit.}
The free energy can be further simplified in the dirty limit
when the impurity scattering rate $\tau^{-1}$ is the largest
among energy scales, apart from the Fermi energy.  In this limit, we can eliminate the momentum integration
and express the energy in terms of the momentum-averaged GF, which we
denote as $g_s = \langle g \rangle$.

The expression which has been used \cite{PhysRevApplied.6.054002,PhysRevLett.121.077002,Eltschka2015,PhysRevB.99.104504}
for the dirty superconductors with spin-singlet
s-wave pairing described by the pairing constant $V$ reads
\begin{align} \label{Eq:FreeEnergyGen0}
  & \frac{F_s}{N_0} =  \frac{|\Delta|^2}{V}
  \\ \nonumber
  &
  -\frac{\pi T}{2} \sum_{\omega }
     {\rm tr}
     \{ (\omega_n + i \vec{h}\cdot\vec{\sigma})\hat{\tau}_3 
     \hat g_s
     + \hat{\Delta}\hat{g}_s  -\frac{D}{4} (\check{\nabla} \hat g_s)^2 \}
\end{align}
The saddle point of this expression yields the Usadel equation for $\hat{g}_s$ and the self-consistency
equation for $\hat{\Delta}$, and therefore \eqref{Eq:FreeEnergyGen0} is naturally considered as the free energy candidate.
A similar expression can also be derived from diffusive nonlinear $\sigma$-models \cite{altland2000-ftm,kamenev2010-kta}.
In order to discuss this result in the Luttinger--Ward framework,
where $\hat{\Delta}$ is handled in a slightly different way,
we need to first substitute in the saddle-point value $|\Delta|^2/V=\frac{1}{4}\Tr\hat{\Delta}\hat{g}$:
\begin{align} \label{Eq:FreeEnergyGen}
  & \frac{F_s}{N_0} =
  -\frac{\pi T}{2} \sum_{\omega }
     {\rm tr}
     \{ (\omega_n + i \vec{h}\cdot\vec{\sigma})\hat{\tau}_3 
     \hat g_s
     + \frac{1}{2} \hat{\Delta}\hat{g}_s  -\frac{D}{4} (\check{\nabla} \hat g_s)^2 \} .
\end{align}
Here we allow arbitrary coordinate dependence of exchange field $\bm h(\bm r)$, the presence of SOC and vector potential in the covariant gradient operator $\check{\nabla}$. 
This expression can be directly derived from Eq.~\eqref{Eq:substitutedF}, by including
the impurity scattering:
The terms without gradients in (\ref{Eq:FreeEnergyGen}) are obtained immediately from the $\Lambda$ and $\Phi$ terms in (\ref{Eq:substitutedF}) by replacing the exact GF with $\hat g_s$. Below we explain how to obtain the gradient terms as well.

Within Born approximation, the impurity scattering can be described by the self-energy 
and the corresponding contribution to the generating functional given by 
\begin{align} \label{Eq:ImpuritySelfEnergy}
  \hat\Sigma_{\rm imp} &= 
  \hat g_s /2\tau
  \,,
 &
 \Phi_{\rm imp} &=
 \Tr (\hat 1 -\hat g_s^2) /8\tau
 \,.
\end{align}
To obtain the free energy functional in the limit $\tau\to0$,
we expand the solution of Eq.~\eqref{eq:glambda-condition} in spherical harmonics,
\begin{align} \label{Eq:GFexpansion}
  \hat g &\approx \hat g_s + {\vec{n}_p}\cdot\hat{\vec{g}}_a
  &
  \hat{\vec{g}}_a &= - l \hat g_s {\check{\nabla}}\hat g_s
  \,,
\end{align}
where $l=v_F \tau$. The 
anisotropic contribution $\hat{\vec{g}}_a$ is determined by  
the Eilenberger equation (\ref{eq:glambda-condition}).

We first evaluate $\Phi_{\rm imp}$:
\begin{align} \label{Eq:JimpDf}
  \Phi_{\rm imp} =
   \frac{ \Tr \hat{\vec{g}}_a^2}{24\tau} = 
  -\frac {D}{8} \Tr ( {\check{\nabla}} \hat g_s)^2,
\end{align}
where $D=v_F{}l/3$ is the diffusion constant.
Here, we noted the normalization condition $\hat g^2=1$ averaged over
directions implies $\hat g^2_s\simeq \hat 1-\hat{\vec{g}}_a^2/3$, where it is now important
to retain the second-order term in $l$.
The last equality follows from
$l(\hat g_s {\check{\nabla}}\hat g_s) (\hat g_s {\check{\nabla}} \hat g_s) \simeq -l( \check{\nabla} \hat g_s)^2$,
which holds in leading order due to the normalization condition.

A similar contribution appears from the gradient term functional $E$ \eqref{eq:Egeneral}.  We can
first observe from Eq.~\eqref{eq:Egeneral} that for matrices
$\hat{g}_s$ without angular dependence, $E[\hat{g}_s]=0$, because of
the angular average in $\Tr$.  
  In the leading order in $l$ the anisotropic correction~\eqref{Eq:GFexpansion} can be considered as a variation of the GF. 
 Then we can calculate the 
 value of the functional $E[\hat g]$
 by using its defining property
(\ref{Eq:ElVarialtion}) :
\begin{align}
  \int d^3r\,
  E[\hat{g}_s + {\vec{n}_p}\cdot\hat{\vec{g}}_a]
  =
  \int d^3r\,
  \Tr \hat{W}(\vec{v}_F\cdot\check{\nabla}\hat{g}_s)
  +
  \mathcal{O}(l^2)
  \,,
\end{align}
where the matrix $\hat W$ is such that
\begin{align}
  [\hat W, \hat g_s] = - l \hat g_s ({\vec{n}_p}\cdot {\check{\nabla}})\hat g_s
  \,.
\end{align}
This implies
\begin{align}
  \hat W (\vec{v}_F\cdot\check{\nabla})\check{g}_s = v_Fl({\vec{n}_p}\cdot\check{\nabla}\check g_s)^2 + \check{g}_s\hat W \check{g}_s \check{\nabla} \check{g}_s
\end{align}
so that, taking into account that $\check{g}_s \check{\nabla}\check{g}_s=-\check{\nabla}\check{g}_s\check{g}_s + \mathcal{O}(l^2)$,
we obtain
\begin{align}
  \Tr \hat W (\vec{v}_F\cdot\check{\nabla})\check{g}_s
  \simeq
  \frac{v_Fl}{2}
  \Tr ({\vec{n}_p}\cdot\check{\nabla}\check{g}_s)^2
  \,.
\end{align}
Then Eq.~\eqref{Eq:ElVarialtion} yields the gradient term
$E \simeq \frac{D}{2}\Tr ({\check \nabla} \hat g_s)^2$,
so that $\frac{1}{2}E + \Phi_{\rm imp} = \frac{D}{8}\Tr({\check \nabla} \hat g_s)^2$.
This leads to the free energy functional in the diffusive limit (\ref{Eq:FreeEnergyGen}). 

{\bf Summary and discussion.}
We have rigorously derived the  free energy functional (\ref{Eq:substitutedF},\ref{eq:Egeneral})
of a superconducting system in terms of the quasiclassical propagators. We obtained convenient expressions in terms of 
Riccati amplitudes (\ref{Eq:RiccatiFreeEnergy}) and in the diffusive limit (\ref{Eq:FreeEnergyGen}).  The functional generalizes the well-known Eilenberger free energy 
for the systems with arbitrary type of pairing and  interacting with spin-dependent fields. 
The result fills an important gap in the theory of superconductivity between the Eilenberger free energy and the Luttinger-Ward 
functional. It can be used to analyze thermodynamic properties of many 
superconducting systems, some of which attract intense interest nowadays.
Among them there are exotic states in unconventional superconductors  \cite{Holmvall2018,PhysRevLett.102.177001,Haakansson2015,PhysRevLett.98.045301} and various hybrid systems \cite{RevModPhys.77.935} including those with spin-triplet superconducting correlations produced either by 
the exchange field and/or SOC \cite{RevModPhys.77.1321, Bergeret2014a}.
Superconductor/ferromagnet systems are studied quite intensively in view of spintronic applications
\cite{Linder2015a, Eschrig2015a}.
With the help of free energy expressions found in this Letter it is possible to analyze complicated behaviour of competing superconducting phases such as $0$-$\pi$
Josephson junctions \cite{RevModPhys.77.935}, 
cryptoferromagnetism 
\cite{Bulaevskii1985,PhysRev.116.898, PhysRevB.62.11872, Buzdin1988,Izyumov:2002},  FFLO states \cite{PhysRev.135.A550,Larkin1965}
modified by different geometrical factors \cite{PhysRevLett.121.077002,PhysRevLett.109.237002} and 
configurations with different vorticity
\cite{barkman2019surface,PhysRevB.79.174514} in such systems using rigorous microscopic calculations.

The interplay of  SOC and external magnetic field
generates proximity-induced topological superconductivity in Majorana nanowires \cite{PhysRevLett.105.177002}. The ground state of such systems 
taking into account the important orbital effect and Abrikosov vortex formation \cite{PhysRevLett.122.187702,PhysRevB.93.235434,PhysRevLett.122.187702}
can be found by calculating the free energy, which can be done using our expressions with arbitrary impurity scattering rate. 

Finally, let us mention the possibility of applying our results to study the free energy of spin-triplet superconductors\cite{RevModPhys.75.657} such as Sr$_2$RuO$_4$ and superfluid $^3$He under various conditions \cite{vollhardt1990superfluid}. Even though the spin-triplet superfluity in $^3$He has been studied for many years, the Eilenberger-type free energy expression is derived only in the present work, which therefore can be considered as a significant advance in the theory of spin-triplet paired states. This tool should be particularly useful to study different competing and spatially-inhomogeneous phases for the confined  topological superfluids \cite{PhysRevLett.98.045301,vorontsov2003-tpt,Levitin2013,PhysRevLett.121.045301}, exotic disordered phases \cite{dmitriev2015polar,PhysRevB.73.060504} and 
vortex states such as double-core vortices \cite{salomaa1987quantized,lounasmaa1999vortices,salomaa1983vortices,thuneberg1986identification,thuneberg1987ginzburg,Fogelstrom1995,Fogelstrom1999,1908.04190} and recently found half-quantum vortices \cite{PhysRevLett.117.255301,makinen2019half}.

\acknowledgments

The work of M.S. and A.V. was supported by the Academy of Finland
(Project No. 297439), and P.V. by the European Union Horizon 2020
research and innovation programme under grant agreement No.~800923
(SUPERTED).

\bibliography{sfs}

\appendix

\input{appendix.tex}

\end{document}

%% file: appendix.tex
\setcounter{equation}{0}
\setcounter{figure}{0}
\renewcommand{\theequation}{S\arabic{equation}}
\renewcommand{\thefigure}{S\arabic{figure}}

\section{Derivation of Eq.~(12)}
\label{app:genvar}

We now derive Eq.~(12).  We assume $\hat{\tau}^2=1$,
$\hat{g}^2=1$, $\partial_\lambda\hat{g}=[\hat{W},\hat{g}]$, and
$\partial_\lambda\hat{\tau}=0$.  Moreover, we denote
$\partial\equiv\frac{\vec{v}_F}{v_F}\cdot\check{\nabla}$ as
the derivative operator in the Eilenberger equation.

From the above, it follows, with standard matrix calculus, $\partial(\hat{a}^{-1})=-\hat{a}^{-1}(\partial a)\hat{a}^{-1}$,
$\delta(\hat{a}^{-1})=-\hat{a}^{-1}(\delta a)\hat{a}^{-1}$, and moreover
$\partial\hat{g} \hat{g} = -\hat{g}\partial\hat{g}$, and
$\delta\hat{g} \hat{g} = -\hat{g}\delta\hat{g}$.
Denote $\hat{Z}\equiv{}[\hat{\tau},\hat{g}]^{-1}$.
We can observe that $\hat{Z}\hat{g}=-\hat{g}\hat{Z}$ and
$\hat{Z}\hat{\tau}=-\hat{\tau}\hat{Z}$.

Equipped with the above, consider then the variation vs. $\hat{g}$
of $E_s=\frac{1}{2}\tr\hat{g}[\hat{\tau},\hat{g}]\partial[\hat{\tau},\hat{g}]^{-1}$,
\begin{align}
  \label{eq:2dE}
  2\delta E_s
  &=
  \delta\tr\hat{g}[\hat{\tau},\hat{g}]\partial Z
  \\
  \notag
  &=
  \tr\delta(\hat{g}[\hat{\tau},\hat{g}])\partial\hat{Z}
  -
  \tr\partial(\hat{g}[\hat{\tau},\hat{g}])\delta\hat{Z}
  +
  \tr\partial(\hat{g}[\hat{\tau},\hat{g}]\delta\hat{Z})
  \\\notag
  &=
  2\delta E_1 - 2\delta E_2 + 2\delta E_3
  \,.
\end{align}
We write $\delta{}E=\delta{}E'+\delta{}E''$ where
$\delta{E}'$ do not contain terms $\propto\partial\hat{\tau}$. We have:
\begin{align}
  2\delta E_1'
  &=
  -
  \tr
  (\delta\hat{g}[\hat{\tau},\hat{g}] + \hat{g}[\hat{\tau},\delta\hat{g}])
  \hat{Z}
  [\hat{\tau}, \partial \hat{g}]
  \hat{Z}
  \\\notag
  &=
  -
  \tr
  \Bigl(
  \delta\hat{g} [\hat{\tau}, \partial \hat{g}] \hat{Z}
  +
  \partial \hat{g}
  \hat{Z}
  (
  \hat{\tau}\hat{g}[\hat{\tau},\delta\hat{g}]
  -
  \hat{g}[\hat{\tau},\delta\hat{g}]\hat{\tau}
  )
  \hat{Z}
  \Bigr)
  \\\notag
  &=
  -
  \tr \delta\hat{g} [\hat{\tau}, \partial \hat{g}] \hat{Z}
  -
  \tr
  \partial\hat{g}
  [\hat{\tau},\delta\hat{g}]
  \hat{Z}
  \\\notag
  &\quad
  -
  2
  \tr
  (\partial \hat{g})
  \hat{Z}
  \hat{g}
  (
  \delta\hat{g}
  -
  \hat{\tau}\delta\hat{g}\hat{\tau}
  )
  \hat{Z}
  \,.
\end{align}
The term $\delta E_2'$ is obtained by exchanging
$\partial$ and $\delta$ in the above expression.
We then find
\begin{align}
  \notag
  \delta E_{1}'-\delta{}E_2'
  &=
  \tr
  \bigl[
  \delta \hat{g}
  \hat{Z}
  \hat{g}
  (
  \partial\hat{g}
  -
  \hat{\tau}\partial\hat{g}\hat{\tau}
  )
  -
  \partial \hat{g}
  \hat{Z}
  \hat{g}
  (
  \delta\hat{g}
  -
  \hat{\tau}\delta\hat{g}\hat{\tau}
  )
  \bigr]
  \hat{Z}
  \\\notag
  &=
  \tr
  \delta \hat{g}
  \hat{Z}
  \Bigl(
  \hat{\tau}
  \hat{g}
  \partial \hat{g}
  \hat{\tau}
  -
  \hat{g}
  \hat{\tau}
  \partial\hat{g}
  \hat{\tau}
  \Bigr)
  \hat{Z}
  \\\notag
  &=
  \tr
  \delta \hat{g}
  \partial\hat{g}
  \hat{\tau}
  \hat{Z}
  =
  \tr
  [\delta W, \hat{g}]
  \partial\hat{g}
  \hat{\tau}
  \hat{Z}
  \\
  &=
  \tr
  (
  \partial\hat{g}
  )
  \delta W
  \,.
\end{align}
Moreover,
\begin{align}
  \notag
  &\delta E_{1}''-\delta E_2''
  \\\notag
  &=
  \frac{1}{2}
  \tr \hat{g}[\partial\hat{\tau},\hat{g}]\hat{Z}[\hat{\tau},\delta\hat{g}]\hat{Z}
  -
  \frac{1}{2}
  \tr \delta(\hat{g}[\hat{\tau},\hat{g}])\hat{Z}[\partial\hat{\tau},\hat{g}]\hat{Z}
  \\\notag
  &=
  \frac{1}{2}
  \tr \partial\hat{\tau}\hat{Z}
  \bigl(
  [\hat{g},\delta(\hat{g}[\hat{\tau},\hat{g}])]
  +
  \hat{g}[\hat{\tau},\delta\hat{g}]\hat{g}-[\hat{\tau},\delta\hat{g}]
  \bigr)\hat{Z}
  \\
  &= 0
  \,.
\end{align}
We then find,
\begin{align}
  \label{eq:Evariationtotal}
  \delta E_s
  =
  \tr
  (
  \partial\hat{g}
  )
  \delta W
  +
  \frac{1}{2}
  \tr\partial(\hat{g}[\hat{\tau},\hat{g}]\delta\hat{Z})
  \,.
\end{align}
The functional~(12) then indeed has the claimed
variation in the interior.  Note that the above calculation did not
assume a specific form for the matrix $\hat{\tau}$.

We can also evaluate the variation vs. $\hat{\tau}$:
\begin{align}
  \notag
  \delta_\tau E_s
  &=
  -
  \frac{1}{2}
  \tr\delta\hat{Z} \hat{g} \partial[\hat{\tau},\hat{g}]
  +
  \frac{1}{2}
  \tr[\delta\tau,\hat{g}]\partial(\hat{Z}\hat{g})
  \\\notag&\qquad
  -
  \frac{1}{2}
  \tr\partial([\delta\tau,\hat{g}]\hat{Z}\hat{g})
  \\\notag
  &=
  \frac{1}{2}
  \tr[\delta\tau,\hat{g}]\hat{Z}
  (\hat{g}\partial[\tau,\hat{g}]+(\partial[\tau,\hat{g}])\hat{g}+(\partial{}g)[\tau,\hat{g}])\hat{Z}
  \\\notag&\qquad
  +
  \tr\partial(\delta\tau\hat{Z})
  \\
  \label{eq:E-tau-variation}
  &=
  \tr\partial(\delta\tau\hat{Z})
  \,,
\end{align}
which is a full derivative.

Integrating Eq.~(8) now reduces to an application
of the Stokes theorem. 
In particular, Eq.~\eqref{eq:Evariationtotal} implies
\begin{align}
  \label{eq:stokes-form}
  \tr
  \partial\hat{g}_\lambda
  W_\lambda
  =
  \partial_\lambda E_s
  -
  \partial_s E_\lambda
  \,,
\end{align}
where
$E_\lambda=\frac{1}{2}\tr(\hat{g}_\lambda[\hat{\tau},\hat{g}_\lambda]\partial_\lambda[\tau,\hat{g}_\lambda]^{-1})$,
and we write
$\partial_s\tr\hat{X}\equiv\vec{n}\cdot\nabla\tr\hat{X}=\vec{n}\cdot\tr\check{\nabla}\hat{X}$.
Hence,
\begin{align}
  &\int_0^1 d\lambda
  \int d^3r\,
  \Tr[v_F\partial \hat{g}_\lambda W_\lambda]
  =
  \Bigl\langle
  \int d^2\rho
  v_F
  \int_{\partial M} d\vec{l}\cdot \vec{E}
  \Bigr\rangle_{\hat{p},\omega}
  \notag
  \\
  &=
  \int d^3r\, (E[\hat{g}_1] - E[\hat{g}_0])
  \\\notag
  &
  \qquad
  +
  \Bigl\langle
  \int d^2\rho
  \int_0^1 d\lambda\, v_F (E_\lambda\rvert_{s=\infty} - E_\lambda\rvert_{s=-\infty})
  \Bigr\rangle_{\hat{p},\omega}
  \,,
\end{align}
where $\langle{X}\rangle{}_{\hat{p},\omega}=\pi T
N_0\sum_{\omega_n}\int\frac{d\Omega_p}{4\pi} X$ so that $\Tr X =
\langle{\tr X}\rangle_{\hat{p},\omega}$.  The line integral is over
the boundary of $M=[0,1]\times[-\infty,\infty]$ with
$d\vec{l}=(d\lambda, ds)$ and $\vec{E}=(E_\lambda,E_s)$.
The spatial integral is decomposed to an integral over the coordinate
$s$ along $\vec{n}$ and the perpendicular coordinate $\vec{\rho}$.

The last boundary term vanishes under the average over momentum
directions, if $v_F(-\hat{p})=v_F(\hat{p})$. It also vanishes if the
boundary conditions for $\hat{g}_\lambda$ are equal,
$\hat{g}_\lambda(s=\infty)=\hat{g}_\lambda(s=-\infty)$,
or if they are independent of $\lambda$ (e.g. normal state at
infinity). This also indicates the boundary term can be neglected when
studying local effects in infinite systems.

We need to observe that the above results assume
$[\hat{\tau},\hat{g}]$ is invertible everywhere in $M$, since
Eq.~\eqref{eq:stokes-form} does not apply at the singularities where
$\vec{E}$ is not defined. Such points give additional contributions
that have to be subtracted, i.e., $\partial M$ includes also clockwise
contours $C_*$ (with infinitesimal interior) circling each
singularity lying inside $[0,1]\times[-\infty,\infty]$.  Each gives a
contribution
\begin{align}
  \oint_{C_*}\tr[gZ^{-1}dZ]
  \,,
\end{align}
Note that because
$\tr[\hat{g}\hat{Z}^{-1}[A,\hat{Z}]]=-2\tr[\hat{g}A]$, gauge fields do
not contribute, and we replaced $\partial\mapsto\partial_s$,
and $dZ=\partial_sZ ds + \partial_\lambda Z d\lambda$.
Writing $g=U\tau_3U^{-1}$, $Z=U\begin{pmatrix}0&w\\\bar{w}&0\end{pmatrix}U^{-1}$ (due to $gZ+Zg=0$),
we have
\begin{align}
  \notag
  \oint_{C_*}\tr[gZ^{-1}dZ]
  &=
  \oint_{C_*}[\frac{1}{2} d\tr(\ln\bar{w}-\ln{}w) - \tr\tau_3U^{-1}dU]
  \\
  &=
  i \pi m
  \,,
\end{align}
where $m$ is an integer. Namely, the last term is regular (we assume
$U$ is nonsingular) and gives no contribution for an infinitesimal
loop, whereas the first terms yield a winding number. The number, and
whether singularities are even present, depends on the choice of
$\hat{\tau}$.  As the free energy is real-valued, these contributions
then can be subtracted by taking the real part.

We find Eq.~(12) indeed gives the bulk contribution to
the derivative term. It is also the only contribution relevant, under
quite general conditions.

\section{Riccati parametrization}

In Ricatti parametrization, the gradient functional can be expressed as
 \begin{align} \label{Eq:EilenbergerFunctionalGeneralSpin}
  & E(\hat g) =
  \\ \nonumber
  & \frac{1}{2}{\rm Tr} [  \bm {v}_F\cdot
  \left( \hat a \nabla \hat b
  - 
  \nabla \hat a \hat b \right)
  (\hat a \hat b)^{-1}
  (1+\hat a \hat b )
  (1-\hat a\hat b)^{-1} ]
  \end{align}
  It is straightforward to check that the variation of this expression by $\hat a$ and $\hat b$ yields gradient terms 
  in the Ricatti equations.  
 This expression can be written in the compact form 
  \begin{align}
    E(\hat g) =  {\rm Tr}  \bm {v}_F\cdot
    [(\nabla \hat a^{-1} + \nabla \hat b) (\hat a^{-1} - \hat b)^{-1}]   + \frac{1}{2}\nabla\ln(\hat a\hat b) ]
 \end{align}
The last term is full derivative and can be neglected.